\documentclass[a4paper,12pt]{article}
\usepackage[utf8]{inputenc}

\usepackage{amsmath,amssymb,theorem}
\usepackage[english]{babel}

\theoremstyle{plain}
\newtheorem{theorem}{Theorem}

\newtheorem{conjecture}[theorem]{Conjecture}
\newtheorem{research_question}[theorem]{Research question}

\begin{document}

\title{Ready for the design of voting rules?}
\author{Sascha Kurz\footnote{
Department of Mathematics, University of Bayreuth, Germany. E-mail: {sascha.kurz@uni-bayreuth.de}.}
}
\date{\today}
\maketitle

\begin{abstract}
  The design of \textit{fair} voting rules has been addressed quite often in the literature. Still, the 
  so-called inverse problem is not entirely resolved. We summarize some achievements in this direction 
  and formulate explicit open questions and conjectures.
  
  \medskip
  
\end{abstract}

\section{Introduction}
Voting in committees with dissimilar committee members raises the question of measuring individual power, i.e., the capability 
to influence a group decision under a given voting rule. To this end, power indices were introduced in order to quantify 
these abilities. Despite the fact that the question for the measurement of power dates back to the late 18th century, at 
the very least, there is still a very active research on power indices. Indeed we believe that this will also be the case 
for the next decades, giving a personal answer to a question about the future of power indices raised by Manfred Holler after 
finishing the collection \cite{30_years_after}. Its predecessor \cite{holler1982power} contains the starting point of a vital 
and still widely open question in the area of power indices. To the best of our knowledge, \cite{nurmi1982problem} was one 
of the first papers addressing the nowadays so-called \textit{inverse (power index) problem}: Suppose one has a rather precise
idea how the power, measured by a specific power index, should be distributed among the committee members -- Penrose's square root law 
may serve as an example. How to design the voting rule (or mechanism) in order to meet the desired power distribution as closely 
as possible?

Of course partial answers have been given during the last 30 years, some of them quite recently, but we think that a lot remains 
to be explored. The aim of this article is to specify this and to propose explicit open questions and challenging conjectures.

\section{Binary voting rules and the inverse problem}
\label{sec_basics}
A lot of models for the description of a voting situation in a committee have been proposed in the literature. Processes of 
opinion-formation, communication structures, a priori unions and dependencies have been considered. The decision space can 
be multi-dimensional, discrete or continuous. Here we want to stick to one of the most simple settings, the binary setting. Given a 
proposal, each voter can either say ``yes'' or ``no'', represented by $1$ and $0$, respectively. The group decision is modeled 
as a simple game $v:2^N\rightarrow\{0,1\}$ mapping the set of supporters to either a ``yes''- or a ``no''-decision (under some 
additional technical requirements). Weighted majority games are one of the most used binary voting rules in practice. Here each voter 
is assigned a non-negative weight $w_i$. If the sum of weights of the supporters meets or exceeds a given quota $q\in\mathbb{R}_{>0}$ 
the group decision is $1$ and $0$ otherwise.

Next we want to formalize the inverse power index problem. Let $\mathcal{C}$ be a class of voting rules, e.g., simple games 
or weighted majority games,  for $n$ voters, $\Vert\cdot\Vert$ be a vector norm in $\mathbb{R}^n$, $\sigma\in\mathbb{R}_{\ge 0}^n$ be a
vector of the desired power distribution, and $P:\mathcal{C}\rightarrow\mathbb{R}_{\ge 0}^n$ be a (positive) power index, e.g., the Shapley-Shubik index 
or the Banzhaf index. With this, the inverse problem asks for a minimizer $v^\star$  of $\Vert P(v)-\sigma \Vert$ over all $v\in\mathcal{C}$.
In practice the determination of an approximate solution $v'$, i.e., where the deviation $\Vert P(v')-\sigma \Vert$ is not too much larger as
the minimum possible value $\Vert P(v^\star)-\sigma \Vert$, is also sufficient. 

Weighted majority games can be \textit{designed} by choosing an appropriate quota and 
appropriate weights. Taking the values of the desired power distribution $\sigma$ as weights $w$ and a quota of  $q=\frac{1}{2}$, 
for example (or some suitable function of the weights), works quite well in some cases, while it does not in others, as we will see 
in the next section.

\section{Limit results}
\label{sec_limit_results}

In its strongest form, the so-called Penrose limit theorem states that, under certain technical conditions, the limit of the fraction 
of the power of two players equals the limit of the fraction of the respective weights. To make things precise, let 
$W\subseteq\mathbb{R}_{\ge 0}$ be a set of weights,
$N^{(0)}\subsetneq N^{(1)}\subsetneq\dots$ be an infinite increasing chain of finite non-empty sets, and $N=\bigcup_{n=0}^{\infty} N^{(n)}$. 
Each voter $i\in N$ is assigned a weight $w_i\in W$. With a fixed real number $q\in(0,1)$, we denote by $\mathcal{W}^{(n)}$ the 
weighted majority game with voter set $N^{(n)}$, weights $w_i$ for all $i\in N^{(n)}$ and quota $q\cdot\sum_{i\in N^{(n)}} w_i$. 
Given a power index $P$ we say that Penrose's limit theorem (PLT) holds for such a chain 
$\left(\mathcal{W}^{(n)}\right)_{n=0}^{\infty}$, if
\begin{equation}
  \label{limit_PLT}
  \lim_{n\to\infty} \frac{P_i\!\left(\mathcal{W}^{(n)}\right)}{P_j\!\left(\mathcal{W}^{(n)}\right)}=\frac{w_i}{w_j}  
\end{equation}
for every pair of voters $i,j\in N$ with $w_i,w_j\neq 0$. We call a player $i$ regular if there exist constants $n_0\in\mathbb{N}$ and 
$\varepsilon>0$ such that 
\begin{equation}
\left|\left\{h\in N^{(n)}\mid w_h=w_i\right\}\right|\cdot \frac{w_i}{\sum_{h\in N^{(n)}} w_h}\ge \varepsilon 
\end{equation}
for all $n\ge n_0$, i.e., the relative weight of all voters having the same weight as voter $i$ does not vanish. We call a chain non-atomic 
(or oceanic), if $\lim\limits_{n\to\infty} w_i/\sum_{h\in N^{(n)}} w_h=0$ for all $i\in N$, i.e., if each individual weight vanishes.

In \cite{lindner2004ls} the authors have proven that PLT holds for regular players in non-atomic chains for the Shapley-Shubik index with arbitrary 
relative quota $q$ and for the Banzhaf index with relative quota $q=\frac{1}{2}$.

\begin{conjecture}
  \label{conjecture_PLT}
  PLT holds for regular players in non-atomic chains for the Banzhaf index with arbitrary 
  relative quota $q\in(0,1)$.
\end{conjecture}

In \cite{kurz2013nucleolus} it is shown that the corresponding statement for the nucleolus as a power index is true.

\begin{research_question}
  For which power indices does Conjecture~\ref{conjecture_PLT} hold?
\end{research_question}

Looking at the example with one voter of weight $1$ and all other voters having weight $2$, we see that, for symmetric power indices, the regularity, of at least the involved voters $i$ and $j$, is necessary. This subtlety, caused by divisibility properties of integers, can be circumvented by replacing
the relative limit~(\ref{limit_PLT}) with
\begin{equation}
  \label{limit_norm_1}
  \lim_{n\to\infty} \sum_{i\in N^{(n)}}\left|  P_i\!\left(\mathcal{W}^{(n)}\right)-w_i\right|=0,  
\end{equation}
i.e., the limit of the norm-$1$-difference between weights and power tends to zero.

Given such a limit result for an arbitrary power index, one can conclude the PLT result for regular players $i$ and $j$ in non-atomic chains. 
In \cite[Proposition 1]{kurz2013nucleolus} the underlying general proof strategy was applied to obtain the PLT result for the nucleolus. 
For the Shapley-Shubik index and non-atomic chains limit equation~(\ref{limit_norm_1}) was shown in \cite{neyman1982renewal}.

\begin{research_question}
  For which power indices does limit equation~(\ref{limit_norm_1}) hold for non-atomic chains?
\end{research_question}

From a practical point of view, the big drawback of those results is the assumption of an infinite chain of weighted majority games. To obtain effective bounds 
for finite games a bit more is needed. And indeed the limit result for the nucleolus $\operatorname{Nuc}$ is based on the inequality
\begin{equation}
  \left\Vert \operatorname{Nuc}([q;w_1,\dots,w_n])-(w_1,\dots,w_n)\right\Vert_1\le \frac{2\Delta}{\min(q,1-q)},
\end{equation}
where $\Delta=\max_{1\le i\le n}w_i$, $q\in(0,1)$, $w\in\mathbb{R}_{\ge 0}$, and $\Vert w\Vert_1=1$, i.e.\ where $\Delta$ is the 
maximum weight assuming a normalized representation. We remark that $\Delta$ tends to zero in the non-atomic case.

\begin{research_question}
  \label{rq_explicit_bounds}
  For which power indices do constants $\alpha,\beta,c\in\mathbb{R}_{>0}$ exist, such that 
  \begin{equation}
    \left\Vert P([q;w_1,\dots,w_n])-(w_1,\dots,w_n)\right\Vert_1\le \frac{c\cdot\Delta^\alpha}{\min(q,1-q)^\beta} 
  \end{equation}
  holds for all $n\in \mathbb{N}$, $q\in(0,1)$, $w\in\mathbb{R}_{\ge 0}$ with $\Vert w\Vert_1=1$ and $\Delta=\max_{1\le i\le n}w_i$?
\end{research_question}

For the Shapley-Shubik index such a bound may implicitly be contained in the technical results of \cite{neyman1982renewal}.

\begin{conjecture}
  The answer to research question~\ref{rq_explicit_bounds} is negative for the Public Good Index.
\end{conjecture}

Of course we are not only interested in an existence result for such constants, but want to know them explicitly (and best possible). 

\begin{conjecture}
  \begin{equation}
    \left\Vert \operatorname{Nuc}([q;w_1,\dots,w_n])-(w_1,\dots,w_n)\right\Vert_1\le \frac{\Delta}{\min(q,1-q)} 
  \end{equation}
  is valid and tight for normalized weights.
\end{conjecture}


So, given 
such a result and a specific desired power distribution $\sigma$, we can compute an upper bound for the norm-$1$-error 
between $\sigma$ as relative weights and the resulting power vector without evaluating the power index. This upper bound 
may also serve as a termination criterion for heuristic search algorithms for the inverse problem.

We can also obtain general assertions about the quality of the solution of taking power as weights in the inverse problem. 
If the maximum weight is not too far away from the average weight, i.e., if there exists a constant $C$ such that relative weights 
are bounded from above by $C\cdot\frac{1}{n}$, then the norm-$1$-error is in $O(\frac{1}{n})$ (for $\alpha=1$), i.e., it decreases at 
least linearly in the number of voters.

The other case, where $\Delta$ does not tend to zero, i.e., if so-called atomic voters with non-vanishing weights are present, 
is also studied in the literature. Given a chain, let us call voters, whose relative weights can be bounded from below by a 
positive constant $\delta>0$, atomic, and those, whose relative weights tend to zero, oceanic. We assume that all voters are either 
atomic or oceanic and that the set $A$ of atomic players is finite. In \cite{shapiro1978values} it is proven 
(cf.~Inequality~4.1) that
\begin{equation}
  \left|\operatorname{SSI}_i\!\left(\mathcal{W}^{(n)}\right)-\operatorname{SSI}_i\!\left(\mathcal{W}^{(m)}\right)\right| \in O\!\left(\frac{1}{n}\right)
\end{equation}
for all atomic players $i$, $n<m$, assuming that the relative weight of the oceanic players does not tend too slowly to zero. A precise 
formula for the limit of the Shapley-Shubik power of an atomic player is stated too. Similar results, while being a bit weaker, have been 
obtained for the Banzhaf index in \cite{dubey1979mathematical}.

\begin{research_question}
  \label{rq_atomic}
  For which power indices does the limit
  \begin{equation}
    \lim_{n\to\infty} P_i\!\left(\mathcal{W}^{(n)}\right)
  \end{equation}
  exist for all atomic players $i\in A$, where $A$ is finite and the relative weights of the non-atomic players 
  can be upper bounded by $C\cdot\frac{1}{n}$ with a constant $C>0$?
\end{research_question}

\begin{conjecture}
  The nucleolus is a positive answer to research question~\ref{rq_atomic}.
\end{conjecture}

\begin{research_question}
  Determine exact formulas for the positive answers to research question~\ref{rq_atomic}.
\end{research_question}

For the nucleolus some first results in that direction are obtained in \cite{galil1974nucleolus}.

\section{Distribution of power vectors}
\label{sec_power_distributions}

Assuming $\sigma\in\mathbb{R}_{\ge 0}^n$ with $\Vert\sigma\Vert_1=1$, the space of the possible 
desired power distributions is an $n-1$-dimensional unit simplex. Since the number of weighted 
majority games or simple games, is finite, not every desired power vector can be met exactly. 
Using asymptotic bounds on the number of games one can conclude a lower bound for a worst case
approximation error, which still tends to zero with increasing $n$, see \cite{kurz2012heuristic}.

Efficient and positive power indices principally map into the $n-1$-dimensional unit simplex, so that
the question arises whether for large $n$ each desired power vector can be approximated suitably well 
by the power vector of a weighted majority or a simple game. Some of the lesser known power indices
satisfy $\frac{1}{c}\cdot\frac{1}{n}\le P_i(v)\le c\cdot\frac{1}{n}$ for all players $i$ and all 
simple games $v$, where $c>0$ is a constant independent from $n$. The Public Help index, see 
\cite{bertini2008public}, is such an example; more can be found in \cite{kurz2014inverse}. As a consequence 
only power distributions, where the entries do not differ too much, can be principally approximated 
well. For the extreme example $(1,0,\dots,0)$ we obtain a best possible approximation error
tending to $2$, which is indeed the maximal distance between two non-negative vectors of sum $1$.

For the Banzhaf index such a constant $c$ does not exist. Nevertheless, e.g., for the power distribution 
$\sigma=(0.75,0.25,0,\dots,0)$ we have $\Vert \operatorname{Bz}(v)-\sigma\Vert\ge \frac{1}{9}$ for 
all simple games $n$, even if $n$ tends to infinity. This example, taken from \cite{kurz2012inverse}, 
is based on the seminal work of \cite{pre05681536}:

\begin{theorem}
  \label{thm_alon_edelman}
  Let $n>k$ be positive integers, $\varepsilon<\frac{1}{k+1}$ be a positive real, and $v$ be a simple game 
  on $n$ voters. If $\sum_{i=k+1}^n \operatorname{Bz}_i(v)\le\varepsilon$, then there exists a simple 
  game $v'$ with $n-k$ dummies such that
  \begin{equation}
    \label{ie_alon_edelman}
    \Vert \operatorname{Bz}(v)-\operatorname{Bz}(v')\Vert_1\le \frac{(2k+1)\varepsilon}{1-(k+1)\varepsilon}+\varepsilon.
  \end{equation}
\end{theorem}

In other words the Banzhaf vector of a simple game, where most of the power is concentrated on just $k$ of the $n$ voters, 
has to be \textit{near}, in an explicit sense, to the Banzhaf vector of another $k$-voter simple game.

In \cite{kurz2014inverse} such inequalities where called Alon-Edelman type results and indeed obtained for some other power 
indices besides the Banzhaf index.

\begin{research_question}
  \label{rq_alon_edelman}
  For which power indices do Alon-Edelman type results do exist?
\end{research_question}

We remark that for the Johnston index, assuming a minor technical condition, no such result can exist.

\begin{conjecture}
  The nucleolus, the minimum sum representation index, the Shapley Shubik index and, more generally,  
  $p$-binomial semivalues are positive answers to research question~\ref{rq_alon_edelman}.
\end{conjecture}

For practical non-approximability computations the concrete bounds are of importance as well. The right
hand side of (\ref{ie_alon_edelman}) could indeed be decreased to $(2k+2)\varepsilon$ in \cite{kurz2014inverse}
and we ask for further improvements and worst case examples.

\begin{conjecture}
  For $\sigma^n=(0.75,0.25,0,\dots,0)\in\mathbb{R}_{\ge 0}^n$ and for each simple game $v$ 
  we have $\Vert \operatorname{Bz}(v)-\sigma^n\Vert_1\ge \frac{14}{37}$ and $\Vert \operatorname{SSI}(v)-\sigma^n\Vert_1\ge \frac{1}{3}$.
\end{conjecture}

\begin{conjecture}
  For each $\sigma\in\mathbb{R}_{\ge 0}^n$ with $\Vert\sigma\Vert_1=1$ and $n\ge 3$ there exists a weighted majority game $v$ with
  $\Vert \operatorname{SSI}(v)-\sigma^n\Vert_1\le \frac{1}{3}$.
\end{conjecture}

What about the nucleolus and other power indices? Are the worst-case examples unique? What happens when we consider only purely 
oceanic games? 

With respect to the latter question, we consider the power distribution $\psi^n=\frac{1}{2n-1}\cdot (2,\dots,2,1)$. In 
\cite{kurz2012heuristic} the inequality
\begin{equation}
  \label{ie_twos_plus_one}
  \Vert P([q;\psi^n])-\psi^n\Vert_1\ge \frac{2}{2n-1}\cdot \frac{n-1}{n}\in\Theta\!\left(\frac{1}{n}\right)
\end{equation}
was shown for all symmetric, positive and efficient power indices $P$ for an arbitrary relative quota $q$. So for this special 
sequence of desired power distributions, taking the desired power vector as weights, yields an $\Omega(1/n)$-error, which is still 
considerably large for medium sized constitutions like, e.g., $n=27$ or $n=28$.

Additionally, the authors solved the inverse power index problem for $\psi^n$ and the Banzhaf index. It turned out that within the 
class of weighted majority games, taking the optimal solution, the lower bound of (\ref{ie_twos_plus_one}) could be decreased by a small
factor only. To the contrary, there exist exact solutions within the class of simple games for $6\le n\le 18$ voters.

\begin{conjecture}
  There exists a constant $c>0$ such that $\Vert \operatorname{Bz}(v)-\psi^n\Vert_1\ge \frac{c}{n}$ for all 
  $n\in\mathbb{N}_{\ge 2}$ and all weighted majority games $v$.
\end{conjecture}

\begin{conjecture}
  For all $n\ge 6$ there exists a simple game $v$ such that $\operatorname{Bz}(v)=\psi^n$.
\end{conjecture}

\begin{research_question}
  Does there exist a sequence of desired power distributions $\sigma^n\in\mathbb{R}_{\ge 0}^n$ with 
  $\Vert \sigma^n\Vert_1=1$ and $\sigma_i^n\in O(1/n)$ for all $1\le i\le n$ such that 
  $\Vert \operatorname{Bz}(v)-\sigma^n\Vert_1\in \Omega(1/n)$ for all weighted majority (or simple) games $v$? 
\end{research_question}

It seems that for the Banzhaf index 
there are some regions near the boundary of the unit simplex which can not be achieved as Banzhaf vectors of simple games.
In the interior of the unit simplex the worst case distance to a Banzhaf vector of a weighted majority game is conjectured to be 
of order $\frac{1}{n}$. 

\begin{research_question}
  Can we obtain more structural results on the set of achievable Banzhaf vectors?
\end{research_question}
To make thinks more precise, we state a few of the known results. Let $\eta_i$ denote the number of swings of player 
$i$, i.e., the Banzhaf vector is proportional to the vector $\eta=(\eta_1,\dots,\eta_n)$ of swings. The minimum and 
the maximum number of total swings, i.e., $n\le \Vert\eta\Vert_1\le \left(\left\lfloor\frac{n}{2}\right\rfloor+1\right)\cdot{n\choose \left\lfloor\frac{n}{2}\right\rfloor+1}$, are well known. Additionally, the $\eta_i$ are either all odd or all even numbers. If the underlying 
game is decisive, then the $\eta_i$ are all even and $\eta_i-\eta_j$ is divisible by $4$, see e.g.\ \cite{dubey1979mathematical}.

Of course the corresponding questions can be stated for other power indices too. 

\section{Algorithmic approaches}
\label{sec_algorithms}
The first algorithms that were proposed for the inverse power index problem were heuristics (mainly hill-climbing and local-search),  
 see e.g.\ \cite{kurz2012heuristic}. Those heuristics are usually quite fast and produce small approximation errors on \textit{practical 
instances}. Given the results from the previous section, a fundamental problem arises. Those methods do not provide a priori 
or a posteriori bounds on the achievable approximation error, which indeed may be large.

Conceptionally the easiest exact and general approach is based on exhaustive enumeration of a given class of voting games. For each 
game the power distribution, according to the chosen power index, and its corresponding distance to the desired power 
distribution can be computed easily. Taking the game with the smallest obtained distance solves the inverse problem exactly. Exhaustively 
generating weighted majority games, or one of its super classes, is not that hard, when the number of voters is rather small. For
the class of so-called complete simple games there is a well known parameterization theorem, so that these games can be generated efficiently, 
see e.g.\ \cite{kurz2013dedekind}. Since for $n\le 7$ only few complete simple games are non-weighted, checking 
complete simple games for weightedness, see e.g.\ \cite{0943.91005} for several such algorithms, is computationally feasible and was indeed
done in the literature. For $n=8$ the fraction of weighted majority games within the class of complete simple games is roughly $\frac{1}{7}$ and roughly 
$\frac{1}{285}$ for $n=9$. Thus more direct algorithms for the enumeration of weighted majority games are needed. Notwithstanding the
recent progress, see \cite{cutting,design,kurz2012minimum}, the number of weighted majority games is known up to $n=9$ 
voters only.

\begin{research_question}
  Determine the number of weighted majority and complete simple games for $n=10$ voters.
\end{research_question}

Due to the super exponential growth of both complete simple and weighted majority games we conjecture that the corresponding numbers 
for $n=11$ will not be determined exactly in the next decade. Since no intermediate lower bounds for approximability are computed, 
the application of exhaustive enumeration algorithms for the inverse problem is rather constrained.

Recently, algorithms for both the Banzhaf and the Shapley-Shubik index have been designed that can achieve a sufficiently accurate
approximation with running polynomial time, in terms of the number of voters times a factor depending on the desired
approximation quality, see \cite{De:2012:NOS:2213977.2214043,o2011chow,de2012inverse}.

\begin{research_question}
  Develop approximation algorithms for the inverse power index problem for other power indices besides the Banzhaf and the Shapley-Shubik 
  index.
\end{research_question}

Besides the tailor-made approaches also integer linear programming techniques have been applied. For the Banzhaf and the Shapley-Shubik index
an ILP formulation was stated in \cite{kurz2012inverse}. Describing an underlying \textit{counting structure} of most of the known 
power indices, more such formulations  were stated in \cite{kurz2014inverse}. 

\begin{research_question}
  \label{rk_ilp}
  Develop a computationally feasible integer linear programming approach for the inverse problem for the nucleolus and the 
  minimum sum representation index.
\end{research_question}

Given the information which coalition of a simple or weighted majority game is winning in an ILP formulation, it is easy to 
state the Shapley-Shubik power of a voter by a precise formula for example. This is different for the two power indices 
mentioned in research question~\ref{rk_ilp}. In principle, the nucleolus can be described as the solution of a single but 
tremendously sized linear program, so that the dual linear program can be used to obtain such a \textit{formula} within an ILP. See 
\cite{freixas2011alpha} for an application of this technique in the context of cooperative game theory. While we did not try this 
approach, we expect it to be computationally infeasible for rather small $n$.

Still the current optimization approaches for the inverse power index problem are limited to, say, $n=15$ voters, which is far beyond 
what can be achieved by exhaustive enumeration but not completely satisfactory.

\begin{research_question}
  Develop practical hybrid algorithms, which can solve the inverse power index problem routinely for up to, say, $30$ voters, 
  while giving reasonable a posteriori non-approximability bounds.
\end{research_question}

\section{Conclusion}
\label{sec_conclusion}

We have argued that a satisfactory solution of the inverse power index problem needs further progress with respect to 
\begin{enumerate}
 \item[(1)] the so-called limit results, which are currently mainly known for a very few power indices only;
 \item[(2)] our understanding of the distribution of the attainable power vectors within the unit simplex;
 \item[(3)] algorithmic solution approaches.
\end{enumerate}
Indeed a lot needs to be done.

Citing several contemporary papers, we have a well justified hope, that there will be further progress in that
direction during the next decades. This adds to the revelation of properties of power indices. However, for the 
application and the understanding of the political process, the question for the most appropriate decision rules
is even more important than the hunt for the \textit{right index}.

In order to encourage research on the broader context of the inverse problem, we have listed several research 
questions and explicit conjectures, which deserve to be either proven or falsified.

To answer the question of the title, I indeed believe that we are ready to design voting games for the inverse problem in
the binary setting, but still more research is needed.


\end{document}